# Mo Thio and Oxo-Thio Molecular Complexes Film as Self-Healing Catalyst for Photocatalytic Hydrogen Evolution on 2D Materials.

Juliana Barros Barbosa,[†] Pierre Louis Taberna,[†] Valerie Bourdon,[¥] Iann C. Gerber,[Ŧ] Romuald Poteau,[Ŧ] Andrea Balocchi,[Ŧ] Xavier Marie,[Ŧ] Jerome Esvan,[†] Pascal Puech,[§] Antoine Barnabé,[†] Lucianna Da Gama Fernandes Vieira,[†,⊥] Ionut-Tudor Moraru,[Ŧ,⅂] Jean Yves Chane-Ching.[†*]

[†]Université de Toulouse, UPS, CNRS, CIRIMAT, 118 Route de Narbonne, F-31062, Toulouse, France

[¥]Université de Toulouse, UPS, Service commun, 118 Route de Narbonne, F-31062, Toulouse, France

[Ŧ]Université de Toulouse, INSA-CNRS-UPS, LPCNO, 135 Av. Rangueil, 31077 Toulouse, France

[§]Université de Toulouse, CNRS, CEMES, 29 rue Jeanne Marvig 31055 Toulouse, France.

[⊥]Universidade Federal de Campina Grande, DEMA/CCT/UFPB, Aprigio Veloso, 882, Campina Grande, Brasil

[⅂]Universitatea Babes-Bolyai, Departamentul de Chimie, str. Kogalniceanu, nr. 1, RO-400084, Romania.

**Abstract**

2D semiconducting nanosheets of Transition Metal Dichalcogenides are attractive materials for solar energy conversion because of their unique absorption properties. Here, we propose Mo thio- and oxo-thio-complexes anchored on 2D p-WSe$_2$ nanosheets for efficient water splitting under visible light irradiation with photocurrent density up to 2.0 mA cm$^{-2}$ at -0.2 V/NHE. Besides developing high electro-catalytic activity, the Mo complexe films were shown to display ability to heal surface defects. We propose that the observed healing of surface defects arises from the strong adsorption on point defects of the 2D WSe$_2$ substrate of Mo complexes such as (MoS$_4$)$^{2-}$, (MoOS$_3$)$^{2-}$, (Mo$_2$S$_6$O$_2$)$^{2-}$ as shown from DFT calculations. In addition to display catalytic and healing effects, the thio-, oxo-thio Mo complexes films were shown to enhance charge carrier separation and migration for the hydrogen evolution reaction, thus representing an example of multicomponent passivation layer exhibiting multiple properties.

## 1-Introduction

Two dimensional transition metal dichalcogenide (2D TMDC) materials [1,2] are building blocks of great interest for solar energy conversion [3-6] because of their unique high light absorption properties [7]. While bulk materials possess large optical absorption which is greater than $10^7$ m$^{-1}$ across the visible range, MoS$_2$ monolayer possessing a thickness of ~ 0.8 nm was shown to absorb more than 10 % of light in the visible region of the spectrum [7,8]. Films of TMDC have thus been considered for various applications including ultrathin, flexible photovoltaic devices [1,2] and photo-electrochemical hydrogen production devices [5,6]. The hydrogen evolution reaction (HER) was previously observed on 2D WSe$_2$ under visible irradiation with Pt as a water reduction co-catalyst [5]. Considerable improvement of the photocurrent was recently achieved with an optimized Pt-Cu co-catalyst combined to defects passivation treatments [6]. Nevertheless, to make viable a large scale development of this technology, it is mandatory among others to replace precious metal catalysts with more earth-abundant materials.

Because of the ultrathin thickness of the 2D materials, small sized catalysts such as single atoms [9] or molecular catalysts are well suited to activate the 2D photo-electrodes. Besides perfect size matching, environmentally benign, low-cost (noble metal free), molecular catalysts are perfectly suited for these 2D materials because their small size facilitates infiltration into the layered 2D materials. Among molecular complexes, hydrogenase and nitrogenase enzymes with active centres consisting of Fe, Ni and Mo are nature effective catalysts for the HER [10,11]. Indeed, MoS$_2$ based nanomaterials including molecular [12-15], nanoparticulates [16,17], nanoflakes [18,19], nanosheets [20,21] and amorphous films [22-24], were previously proposed as HER catalysts. Although MoSx amorphous film catalysts displayed thicknesses around 40 -150 nm [24], molecularly thin films are best suited for photocatalytic applications since they minimize light shielding. Molecular Mo complexes including Mo$_3$S$_4$ [12], (Mo$_2$S$_{12}$)$^{2-}$ [14], (Mo$_3$S$_{13}$)$^{2-}$ [13], were recently developed as catalysts for the HER because they represent earth abundant alternatives for large-scale use in place of noble metal catalysts. While very few works have been reported for the catalytic activity of oxo thio complexes [25-27], Mo thio clusters yielding high catalytic efficiency for the HER have been designed to molecularly mimic the MoS$_2$ edge sites. In addition, these thio -Mo- complexes have proven to display reasonable HER stability in acidic solution [13,14]. Thus, these complexes represent ideal candidates to be coated onto 2D photoactive materials such as 2D TMDC.

Synthetic or processing routes of these complexes were largely developed in solvents such as dichloromethane [12], methanol [13] or DMF [14]. In the context of sustainable generation of hydrogen, it is highly desirable to perform the synthesis and to process the Mo complexes in water and open air. Development of a precious metal-free catalyst, stable and active in aqueous environments [28] for the HER has been recently identified as a long term goal. One of the challenges to face during all-aqueous synthesis and film forming is the selection of photoactive species from a species distribution formed at a given pH, as a consequence of the numerous equilibrium reactions usually occurring in $H_2O$. Moreover, due to hydrolysis which occurs inevitably in highly alkaline $H_2O$ medium, the resulting complexes distributions include both thio and oxo-thio-complexes [29]. While catalytic cycles for the HER on thio Mo complexes or clusters were largely documented [12-15], very few studies [25-27] report on the efficiency of oxo-thio groups on HER. Although the catalytic mechanism reported for the Mo=O groups in MoSx amorphous films catalysts [26] involves some redox reactions occurring on the unsaturated Mo sites (Mo(V) + e = Mo(IV)) [30] and thus requiring applied cathodic voltage, recent work [26] argues that the redox reactions of the HER mechanism involve $S_2^{2-}$ anions. Assessment of the photocatalytic activity of these Mo thio complexes for the hydrogen evolution reaction particularly when operating without any applied cathodic voltage should therefore contribute to shed a light onto the involved catalytic mechanism.

Increase of durability of complex catalysts minimizing their catalytic deactivation represents another crucial challenge for the large scale development of these catalysts. Chemical bonding implementation between the complex catalyst and the photoelectrode substrate can prevent complexe desorption[31]. One possible approach is to anchor Mo complexes on the $WSe_2$ photo-electrode via ligand interactions. While internal edge defects such as tears and pinholes can serve as anchoring sites onto the 2D materials [32], this strategy may involve some terminal chemical units of the Mo complexes. This will require a specific design of the Mo complexes in order to preserve the active sites density, ensuring their long term catalytic efficiency. Lastly, another important challenge to face is the defect passivation [5,33] when promoting these 2D nanosheets photocathodes. Defect healing greatly helps to decrease photo-generated charge carriers recombination thus enhancing performance of the photo-electrodes. It is therefore of great interest to explore innovative passivation routes, involving surface defects of the 2D materials interacting with terminal sulfides, or, di-sulfides units of the thio and oxo-thio Mo complexes.

In this work, we propose new water-stable Mo thio complexes as photocatalysts for the hydrogen evolution reaction (HER) in water splitting under visible irradiation. In the context of low cost fabrication, $WSe_2$ electrodes were fabricated by a simple process route including deposition of solvent-exfoliated $WSe_2$ nanosheets by drop casting onto an FTO substrate. The Mo thio complexes were then impregnated onto the $WSe_2$ electrodes via a solvent-free film forming process. Anchoring of the Mo-based molecular catalyst is achieved by selection of the most interactive thio-, oxo-thio- Mo complexes. This involves ligand solution interactions between W(IV) cation or surface defects of the photo-electrode and $S^{2-}$ or $S_2^{2-}$ anions of the thio-complexes catalyst. Resulting films are composed of a mixture of thio and oxo-thio Mo complexes, these latter being inevitably formed in alkaline aqueous medium. A catalyst film structure is proposed mainly resulting from i) identification of the Mo complexes in solution by mass spectroscopy (ESI-MS), ii) collection of XPS data from the resulting catalyst solid film and iii) Density Functional Theory (DFT) calculations of the probable conformations of the Mo thio and oxo thio Mo complexes anchored on $WSe_2$ photo-electrodes. The unique characteristics of thio and oxo thio Mo complexes as healing, anchorable catalysts for the photocatalytic HER on 2D materials are detailed herein. These experimental observations are supported by density functional theory calculations providing insights into (i) the probable binding developed between the surface defects of the 2D materials and the Mo thio and oxo-thio complex catalysts; (ii) the thermodynamics of H* adsorption and $H_2$ desorption on the thio, oxo-thio complexes by means of HER free energy $\Delta G_{H*}$ calculations as proposed by Nørskov *et al.* [34].

**2- Experimental section**

*2-1- Mo thio and oxo thio complexes aqueous solutions preparation*

Ammonium tetrathiomolybdate (($NH_4)_2MoS_4$, Sigma Aldrich) (0.65 g, 2.5 mmole) was dissolved in 0.31 M aqueous ammonium sulfide solution (prepared from $(NH_4)_2S$ 20% wt in $H_2O$, Sigma Aldrich) (8 ml, 2.5 mmole). Subsequent polycondensation of the species is performed by pH adjustment (pH 9.00) by addition of 0.5 M HCl (4.5 ml) and the solution is volume at 25 ml. After aging at 25 °C during 2h, the solution is diluted to 250 ml by $H_2O$. The molar ratio of $S^{2-}$ to Mo, S/Mo is 5, and the Mo concentration is Mo= 0.01M.

*2-2-2D $WSe_2$ – FTO electrode fabrication*

As-received $WSe_2$ powder ($WSe_2$ 99.8% metal basis from Alfa Aesar) was exfoliated in dichlorobenzene solvent (DCB, Aldrich) using ultrasonicator 750 W (Biobloc Scientific). 3 g of $WSe_2$ was exfoliated at 40 % amplitude (6 sec: on / 4 sec: off) for 16 h in a 4 °C bath.

A simple procedure was used to fabricate the $WSe_2$ films. Fluoride doped Tin Oxide (F: $SnO_2$ or FTO) coated glass substrates (SOLEMS, France) were coated by drop casting by a DMF dispersion of $WSe_2$ nanosheets (7.5 g $l^{-1}$ $WSe_2$). After drying by air evaporation at room temperature, 1.5 - 2 g cm$^{-2}$ of $WSe_2$ were deposited after 3 successive coatings.

*2-3-Co-catalyst deposition by selective dip coating impregnation*

Selective dip-coating of the co-catalyst was performed by vertically immersing the $WSe_2$-FTO electrode into the Mo thio complex co-catalyst solution. The electrode was in contact with the co-catalyst solution for 6 h (or 16 h) under gentle stirring at room temperature. The electrode was then washed/rinsed by dropping the electrode into deionized water without stirring for 5 min. A typical sequence includes the three following steps:

1) 16 h impregnation– 5 min washing- 10 min heat treatment at 110 °C

2) 6 h impregnation – 5 min washing- 10 min heat treatment at 110 °C
3) 6 h impregnation – 5 min washing- 10 min heat treatment at 110 °C

*2-4-Hydrogen measurements using Gas Chromatography.*

A closed photo-electrochemical cell equipped with a three electrode set-up was used for the hydrogen detection. An argon flow rate was injected into the cell in order to transport the evolved hydrogen from the photo-electrochemical cell to the Gas Chromatograph (Shimadzu, GC-2014 AT) allowing the real-time analysis of the composition of the gas generated inside the cell.

**3- Results and Discussion**

*3-1-Synthesis and Characterization of a Large Variety of Water-Soluble Thio and Oxo Thio Mo Complexes Distributions.*

In aqueous solutions, Mo(VI) is reduced by $S^{2-}$ with formation of sulfur-rich, polynuclear, di-sulfido complexes [35]. Using $(MoS_4)^{2-}$ as a Mo(VI) source salt and $H_2O$ as solvent, we have explored a large range of water-soluble, thio and oxo-thio complex distributions by variation of polycondensation degree (from pH 8 to pH 9.7) and of $S^{2-}$ concentrations ($S^{2-}/Mo_{mole}$ or $S/Mo_{mole}$ = 4, 5 and 6, S and Mo denote respectively the total concentrations of sulfides and Molydenum species). As a result of the various oxo reduction reactions involved in aqueous sulfide solutions, (Mo(VI) + 2 e$^-$ → Mo(IV) , Mo(VI) + Mo(IV) -> 2 Mo(V) ), the synthesized Mo thio and oxo-thio complexes solutions displayed Mo cations with various oxidation states including Mo(VI), Mo(V) and Mo(IV). Species formed in each solution were rationally identified using Electro-Spray Ionization Mass Spectroscopy (ESI-MS). A typical species distribution diagram for solutions prepared for a total Mo concentration of 0.01 M, pH 9.0 and S/Mo= 5 is reported Fig. 1a. The complexes distribution mainly includes Mo monomers, dimers and in lesser proportion trimers. Using basic aqueous solutions, hydrolysis is favoured and corresponding hydrolysis compounds of the thio complexes were clearly identified in significant proportions. Thus, previously largely described Mo sulfides compounds and their corresponding hydrolysis products such as $(MoS_4)^{2-}$, $(MoS_5)^{2-}$ [36] $(Mo_2S_7O)^{2-}$ [37] and $(Mo_3S_8O)^{2-}$ [38] were unambiguously identified in our solutions (Fig. S1). Interestingly, we also demonstrate from our process route the formation of $(Mo_2S_6O_2)^{2-}$ [35,39], $(Mo_2S_{12})^{2-}$ [40], $(Mo_2S_8)^{2-}$ [41] complexes (Fig. 1a). While these $(Mo_2S_6O_2)^{2-}$, $(Mo_2S_{12})^{2-}$, $(Mo_2S_8)^{2-}$ complexes were largely synthesized in various solvents such as acetonitrile (ACN) or di-methyl-formamide (DMF), our results show the formation of these complexes in $H_2O$ solvent in significant proportions.

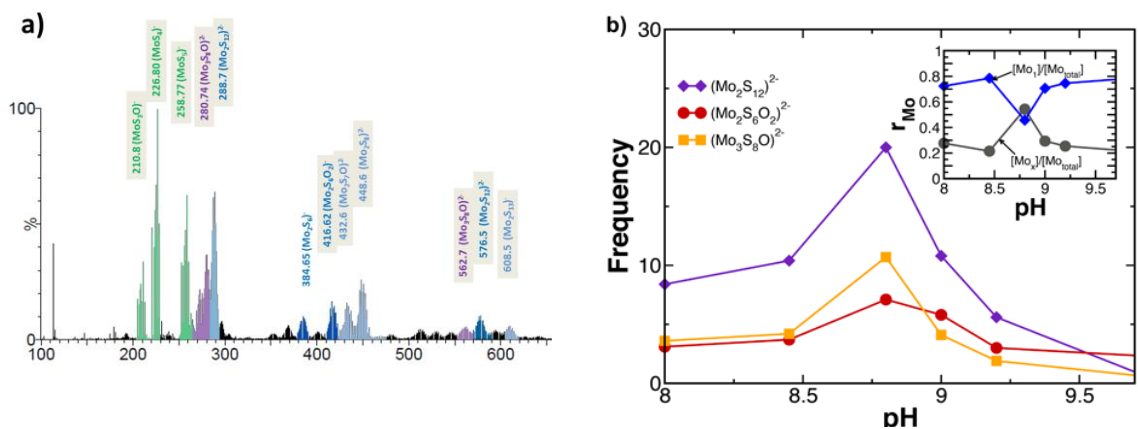

**Fig. 1.** (a) Typical ESI-MS spectra recorded on Mo thio complexes solution prepared at pH 9.00 and S/Mo=5. (b) Species predominance diagram determined at 0.01 M Mo and S/Mo= 5 from ESI-MS data showing concentration peaks for $(Mo_2S_{12})^{2-}$, $(Mo_2S_6O_2)^{2-}$ and $(Mo_3S_8O)^{2-}$ complexes between pH 8.5 and pH 9. Insert: Dependence of $[Mo_1]/[Mo_{total}]$ and $[Mo_x]/[Mo_{total}]$ with pH. $[Mo_1]/[Mo_{total}]$ and $[Mo_x]/[Mo_{total}]$ denote respectively the ratio of Mo monomers, $[Mo_1]/[Mo_{total}]$= Mo monomers/( Mo monomers + Mo polymers) and the ratio of Mo polymers, $[Mo_x]/[Mo_{total}]$= Mo polymers / ( Mo monomers + Mo polymers). Mo monomers detected by ESI MS mainly include $(MoS_4)^{2-}$, $(MoOS_3)^{2-}$

Although peak intensities extracted from the ESI-MS data can depend on ionization properties of the complex, a comparison of the frequencies for a same species at different pH can be achieved quantitatively, allowing determining its predominance domain. From these ESI-MS data, predominance diagram of the various species was defined as a function of pH for each given value of S/Mo, $4 \leq S/Mo \leq 6$. Fig. 1b shows the predominance diagram for Mo sulfides and oxo sulfides species determined for S/Mo= 5. Similarly to others metallic cations [42,43], speciation of the Mo cation clearly shows the coexistence of various species at a given pH. Because increase of pH favours sulfide complexes depolymerisation, Mo sulfide monomers were detected in large proportions in the high pH range. Consistently, formation of polycondensed species occurs at lower pH, $8.75 \leq pH \leq 9.25$. In this pH window, peak concentrations of $(Mo_2S_6O_2)^{2-}$, $(Mo_2S_{12})^{2-}$ dimers and $(Mo_3S_8O)^{2-}$ trimers were clearly observed.

Insights into chemical compositions of the films stemming from these complexes solutions were gained by X-ray photoelectron spectroscopy (XPS). Indeed, characteristics of the whole film can be obtained on films prepared with an ultrathin thickness, e < 10 nm. Various ratios such as the terminal sulfide ratio, $r_{S-ter}$, ($r_{S-ter} = S_{ter} / (S_{ter} + S_{br})$, $S_{ter}$ and $S_{br}$ denote respectively terminal and bridging sulfides), or the Mo oxysulfide ratio, $r_{MoOS}$, ($r_{MoOS} = MoO_yS_z/(MoS_x + MoO_yS_z)_{molar})$) were extracted from XPS data. On films prepared by solution-deposition and subsequent evaporation (or drop casted) on FTO substrates, the higher terminal sulfide ratio (Fig. S2) was observed on the film prepared from the more depolymerized complex solution at pH 9.7. Likewise, Mo oxo-thio complexes identified by ESI-MS on solutions were clearly observed after their evaporation under Ar on the corresponding drop-casted films. Note that the $r_{MoOS}$ maximum value was observed for the film prepared from the solution (S/Mo= 5, pH 8.75-9.0) possessing the highest oxysulfide concentration (Fig. S2) confirming assignment of the intermediate peak observed at $BE_{3d\ 5/2}$ = 230.1 eV to the Mo oxysulfide complexes previously identified in solution by ESI-MS.

*3-2-Electrocatalytic Properties of the thio-, oxo-thio-Mo Complexes Distributions.*

Electro-catalytic properties of films prepared from a large variety of complexes solutions (Mo= 0.01 M, 4 ≤ S/Mo ≤ 6, 8.0 ≤ pH ≤ 9.7) were first explored on FTO substrate. The whole set of water-soluble Mo thio and oxo-thio complexes, pH 8.0 < pH < pH 9.7, including solutions mainly composed of Mo monomers was shown to exhibit reasonably good electrocatalytic properties when deposited on FTO substrate (Fig. S3) with an average overpotential of 260 mV at j= 10 mA cm$^{-2}$. Finer inspection of the current density – potential curves in the low cathodic potential range, -0.2 V < E vs NHE < +0.0 V, reveals slightly better electrocatalytic properties of the films when prepared from complexes solution in the pH window, pH 8.8 ≤ pH ≤ pH 9.2 (Fig. S3). These slightly better catalytic properties are achieved from solutions made up of predominant species composed of Mo monomers $(MoS_4)^{2-}$, $(MoS_3O)^{2-}$ and poly-condensed species possessing two or three Mo centres including $(Mo_2S_6O_2)^{2-}$, $(Mo_2S_{12})^{2-}$ and $(Mo_3S_8O)^{2-}$ clearly identified by ESI-MS. In contrast, slightly lower electrocatalytic properties were observed on films prepared from high pH solutions mainly composed of monomers such as $(MoS_4)^{2-}$, $(MoS_3O)^{2-}$. To clarify the respective contributions on the catalytic performances of the thio and oxo-thio complexes, DMF dispersions of pure $(Mo_2S_{12})^{2-}$ complexes were prepared following a procedure previously described [14]. Properties of these pure $(Mo_2S_{12})^{2-}$ complexes were investigated on films deposited on FTO substrates by drop casting. Let us recall that a value of η = 175 mV at j= 10 mA cm$^{-2}$ was previously reported in the literature [14] for the $(Mo_2S_{12})^{2-}$ thio complex in addition with graphene. Without any graphene addition and in our experimental conditions, i.e. on WSe$_2$ electrodes, similar over-potential values were recorded at j= 10 mA cm$^{-2}$ in a large range of mass loadings for the pure $(Mo_2S_{12})^{2-}$ thio complex as well as for our thio, oxo-thio Mo catalysts (Fig. S4). Thus, our results show that films prepared from our thio-, oxo-thio- complexes display electrocatalytic properties for the HER similar to the $(Mo_2S_{12})^{2-}$ thio complex. Although recent works on single-atom catalysts have shown superior activity, we can note that our Mo-thio, oxo-thio complexes possess catalytic activity in the range of the best Mo-S based catalysts.

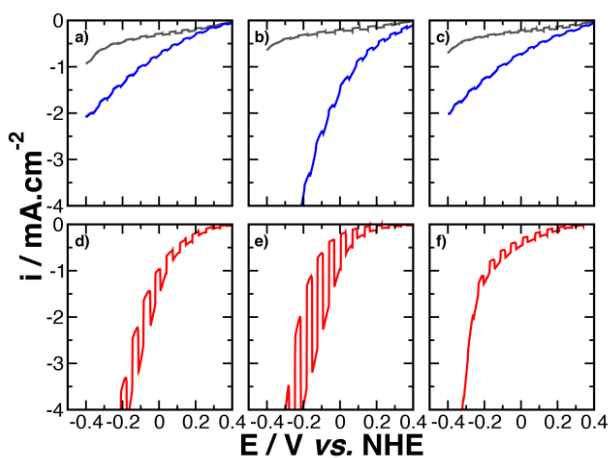

**Fig. 2.** Electrochemical and photo-electrochemical characterization of thio-, oxo-thio- Mo complexes films prepared at S/Mo= 5 and at various pH in $H_2SO_4$ 0.5 M. (Top a-c): J-E curves ($a_1$ pH 8; $a_2$ pH 9; $a_3$ pH 9.5) recorded under intermittent illumination on films prepared by evaporation-drying (or drop casting) onto $WSe_2$ electrodes (blue curves). J-E curves of corresponding films prepared on bare $WSe_2$ electrodes are given as control (black curves). (Down d-f): J-E curves on films prepared by selective dip coating on $WSe_2$ electrodes, impregnation time: 2 x 16 h. The electro-activities of the films can be tracked from the curve sections recorded in the dark.

Using solvent-exfoliated, p-type $WSe_2$ nanosheets, a significant decrease of the electrocatalytic activity is observed for films formed by deposition- evaporation or drop casting onto $WSe_2$ (Fig. 2) substrates compared to films deposited onto FTO (Fig. S3). More specifically, this decrease is clearly observed in the mass loading range of 125-500 nanomoles $cm^{-2}$ for complex solutions displaying a higher content of monomers (pH > 9.25) or prepared in the lower pH range (pH < 8.5). On films drop casted onto $WSe_2$ electrodes, the best electrocatalytic activities were recorded on electrodes formed from solutions prepared in the pH range, 8.5 < pH < 9.25. As previously reported, this pH range corresponds to the predominance domain of Mo dimers or trimers. XPS data performed on the corresponding dried films indicate a significant decrease of ~ 30 % of the terminal sulfide ratio, $r_{S-ter}$, compared with films drop casted on FTO (Fig. 3a). This reveals that the observed decrease of $r_{S-ter}$ probably arises from ligand interactions between terminal $S^{2-}$ or $S_2^{2-}$ anions of the thio or (oxo) thio complexes and W(IV) cations or surface defects of the 2D $WSe_2$ substrate. This ligand interaction involving terminal $S^{2-}$ or $S_2^{2-}$ thus drastically decreases the available $S^{2-}$ or $S_2^{2-}$ catalytic sites concentration of the films prepared from high pH dispersions accounting for the lower catalytic activity of films mainly composed of monomers. In contrast, the relatively higher concentration of $(Mo_2S_6O_2)^{2-}$, $(Mo_2S_{12})^{2-}$ and $(Mo_3S_8O)^{2-}$ complexes allow to preserve a reasonably high concentration of active catalytic sites in the optimal pH range.

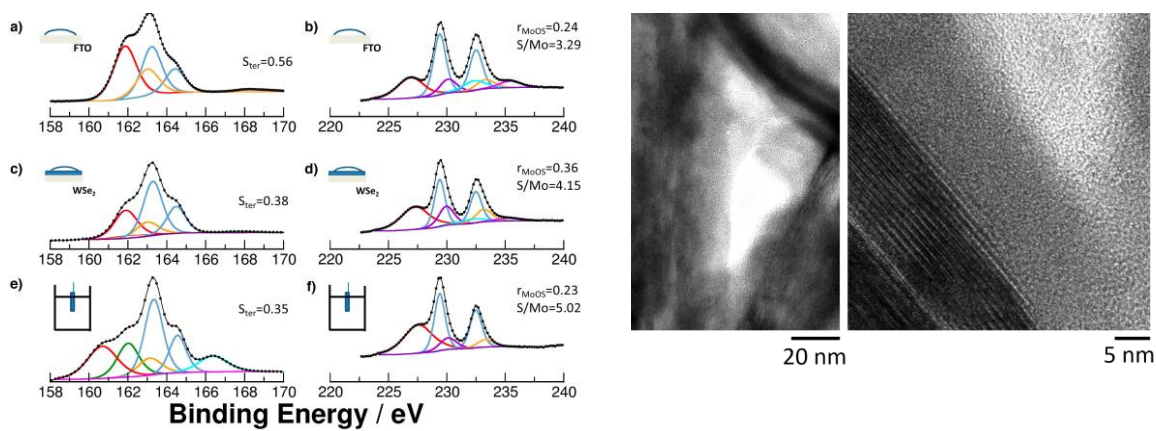

**Fig. 3.** Properties of the thio-, oxo-thio-Mo complexe co-catalyst film. (a) XPS core-level spectra and deconvolution into S2p and Mo 3d contributions for Mo thio complexes thin films deposited by (a-b) drop-casting onto FTO electrodes (c-d) Drop casting onto WSe$_2$ electrodes (e-f) Selective dip coating onto WSe$_2$ electrodes. XPS spectra were recorded after heat treatment 110 °C, Air 10 min, immersion in H$_2$SO$_4$ 0.5 M, 15 min. For the dip coated WSe$_2$ electrodes, XPS spectra were recorded after electrochemical testing. The large decrease of S$_{ter}$ values observed between films deposited on WSe$_2$ compared with FTO electrodes probably arises both from higher interactions (or chemical reactions) between sulfide-disulfide terminal units of the Mo complexes with the WSe$_2$ electrodes. (b) FIB cross-section TEM images showing presence of an ultrathin amorphous coating identified by EDS to the Mo complexe co-catalyst on a well-crystallized anisotropic WSe$_2$ nanoflake. Bright domains represent pores. The high magnification TEM image clearly shows no degradation of the WSe$_2$ at the WSe$_2$- co catalyst interface.

*3-3-Thio- oxo-thio-Mo Complexes as Healing Catalysts for Solar to Hydrogen Conversion.*

The photo-electrochemical performances of the Mo thio complexes -WSe$_2$ electrodes were first assessed on electrodes prepared by drop casting under intermittent illumination (1 sun, 100 mW cm$^{-2}$). Note that all the species included in the MoxSy complexe solution are deposited by evaporation-drying in this film forming process. Low photocurrents were recorded by linear scanning voltammetry (LSV) for the whole set of films prepared in a large range of catalyst loading, 30 nanomoles cm$^{-2}$ < n < 200 nanomoles cm$^{-2}$ (Fig. 2). In spite of the high electro-activity of the MoxSy catalyst films (Fig. S4), the low photocurrent detected suggests a large charge carrier recombination, probably arising from the large concentration of surface defects usually reported for the 2D materials [33,44-45].

In view of the large variety of these Mo complexes distribution solutions, each of which itself being composed of several thio and oxo thio complexes, an interesting approach would be to select the most interactive species taking benefit of the "WSe$_2$-thio complexes" interactions developed in solution. Thus, we have focused on a film forming process favouring the anchorage of thio- and oxo-thio-Mo complexes developing high complexing interactions and high catalytic performances. Similar to a strategy previously proposed to enhance the long term stability of (molybdenum sulfide clusters-defective graphene) catalyst involving S-C covalent bonding [31], chemical bond formation between thio or oxo-thio complexes and WSe$_2$ surface should contribute to improve anchoring of the thio-complexe catalysts onto the photo-electrode. More importantly, strong bonding formation between the thio complexe ligands and surface defects of WSe$_2$ 2D material may help to heal these defects thus improving the optoelectronic performances of the 2D materials. A film forming process (denoted afterwards selective dip-coating) was thus explored involving successive dip coatings with long duration time (up to 2x 16 h) and followed by H$_2$O washing. Interestingly, for dispersions displaying a well-defined range of S/Mo values and polycondensation ratios (or pH), much larger photocurrents were recorded on dip-coated photo-electrodes compared with drop casted films in the large range of film thickness investigated, 50 nmoles cm$^{-2}$ ≤ n ≤ 750 nmoles cm$^{-2}$ (Fig. 2). In contrast to the film prepared by evaporation (or drop casting) which involves the whole set of thio complexes species present in the solution, the better performances of the dip coated films were achieved from a selection of the most catalytically active complexes, mainly driven by the "ligand – WSe$_2$ photoelectrode" interactions acting in the solution during the dip coating process. Several results demonstrate this complexes selection. First, significant differences are observed between the films prepared by drop-casting on FTO vs dip-coating on WSe$_2$ substrates. Optical and electronic microscopy images show formation of dendrite-like acicular particles (Fig. S5a) only visible on FTO, suggesting a phase separation during this film forming process by evaporation-drying, not observed by dip coating. In addition, the main Raman peak recorded on these acicular particles at 400 cm$^{-1}$ (Fig. S5c) is not observed on films spectra prepared on WSe$_2$ photo-electrodes. Second, a slow and progressive increase of the photocatalytic activity is observed with increased time duration clearly recorded throughout the first 32 hours of dip coating (Fig. S6). For the film withdrawn after the first hours of dip coating (t < 4h), XPS data recorded on the corresponding films highlight a peak concentration of the oxo-thio-complexes ($r_{MoOS}$= (MoOS)/(MoOS+MS)$_{molar}$, $r_{MoOS}$= 0.40), probably arising from preferred deposition of (MoO$_y$S$_z$)$^{t-}$ monomers present in larger proportion in the solution. At longer dip coating duration times, changes in the film composition are clearly observed, illustrated at once by a decrease of the Mo-oxysulfide concentration ($r_{MoOS}$= 0.31) and an increase of S/Mo ratio (from S/Mo= 4.2 to 4.6) indicating selection of complexes at high S/Mo ratio. Various driving forces may cause the changes observed in the film composition. In the first stage of dip coating, selection of the thio, oxo-thio-Mo complexes may be dictated by adsorption energies, or, complexation stability constants implying the thio-Mo complexes and the surface entities of the bare $WSe_2$ photo-electrodes. In a second step, the film formation process involves change in the assembly pathway based on interactions between the thio-, oxo-thio- Mo complexes and the first layers of Mo sulfide oligomers anchored on the $WSe_2$ photo-electrode surface. In addition, consistently with this selection process, the best photocatalytic activities were achieved with dip coatings alternated with desorption stages, favouring removal of less interactive species before subsequent impregnation. Lastly, significant differences both of S/Mo and $r_{S-ter}$ ratios are observed on final film compositions between drop casted and dip coated Mo complexes-coated $WSe_2$ electrodes. More particularly, Fig. 3a reveals a selection of high S/Mo ratio complexes including probably a high content of $(Mo_2S_{12})^{2-}$ complexes in the films achieved by selective dip coating vs drop casting process.

From a rationale screening of the photocatalytic activities performed on a large set of films (4 ≤ S/Mo ≤ 7 and 8.0 ≤ pH ≤ 9.7), best photocatalytic performances were observed for dip coated films fabricated from solutions at S/Mo= 5 and in the optimum pH range, 8.75 ≤ pH ≤ 9.25. More interesting, without the use of any passivating additive, much higher photocurrents were achieved after impregnation of the Mo thio complexes catalysts compared with noble metal catalyst such as Pt-Cu [6]. Although the two catalysts exhibit similar electro-activity as revealed from the dark currents, a larger photocurrent of 2.5 mA cm$^{-2}$ is recorded onto the $WSe_2$ photocathode after Mo thio complexes activation vs 0.4 mA cm$^{-2}$, when activated by Pt-Cu (Fig. 4a). Because Pt-Cu catalyst is well known to greatly facilitate electron transfer [6], the large photocurrent improvement observed with Mo thio complexes could be assigned to a drastic decrease of recombination centres density.

In addition, while photocurrent recorded from bare $WSe_2$ photo-electrode exhibits current spikes typical of charge recombination and/or accumulation (positive spikes), such behaviour is completely inhibited for co-catalyst coated $WSe_2$ photo-electrodes (Fig. 4b). We can also notice a similar slope of the i(v) curves recorded under dark or under illumination for the bare $WSe_2$ photo-electrode. In contrast for co-catalyst coated $WSe_2$ photo-electrode, a different slope is clearly observed on the i(v) curve sections recorded under dark or under illumination. This slope increase is likely related to a decrease of recombination rate together with an activation of the charge transfer in line with the healing properties of the co catalyst film. An additional evidence for defects passivation was obtained from photoluminescence. Room temperature photoluminescence (PL) characterization was performed on two $WSe_2$ photo-electrodes with and without Mo thio-complexes catalyst impregnation. While no photoluminescence was recorded on the catalyst-free sample, we clearly observe a peak of the PL at 1.43 eV on the sample coated by the Mo-thio-complex (Fig. 4c).

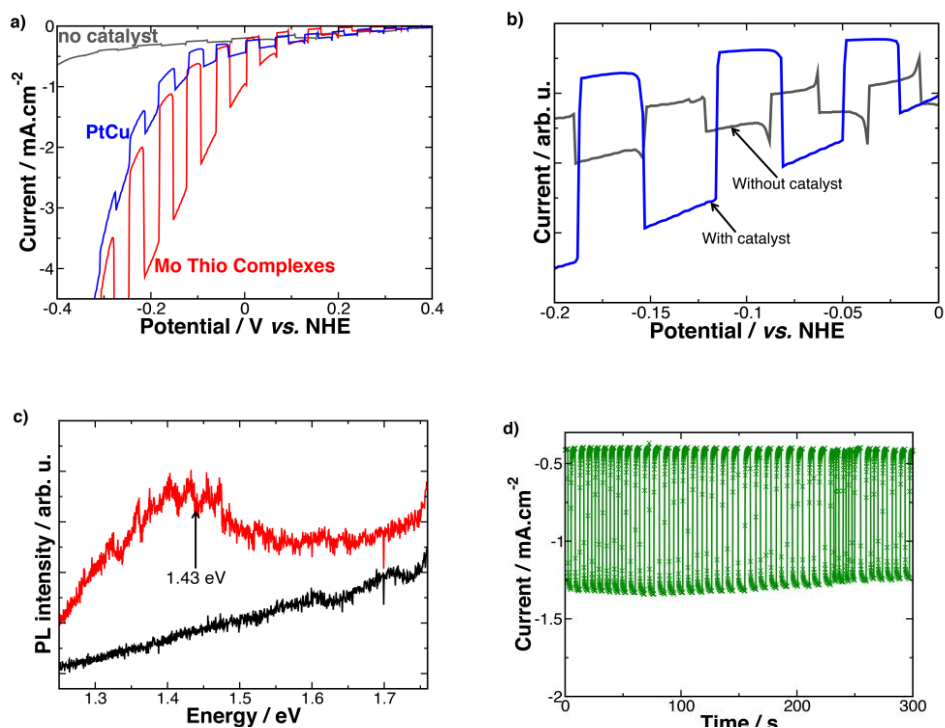

**Fig. 4.** Photo-electrochemical characterization of Mo thio complexes modified WSe$_2$ electrodes. Mo thio complexes are deposited by selective dip coating from complexes distributions prepared at various pH. (a) LSV curves under intermittent (1 sun) illumination showing higher photocurrent on WSe$_2$ photo-electrode after deposition of oxo-thio Mo complexes (S/Mo= 5. 2 x 16 h) compared with HTS free- WSe$_2$ photo- electrode activated by a Pt-Cu catalyst. (b) LSV curves under intermittent illumination showing a significant decrease of transient photocurrent on thio, oxo-thio Mo complexes coated WSe$_2$ photo-electrode compared with a bare photo-electrode. (c) Room temperature photoluminescence (PL) of a WSe$_2$ photo-electrode coated with thio-, oxo-thio-Mo complexes catalyst and prepared from solvent-exfoliated WSe$_2$ nanosheets. The PL recorded on a bare WSe$_2$ photo-electrode without any co-catalyst deposition is given for comparison (Black curve). The photoluminescence (PL) was performed exciting the sample with the red line of a He-Ne laser (E= 1.96 V) and collecting the emitted PL in back-scattering geometry. d) Chrono-amperometric curve recorded on WSe$_2$ electrodes activated by Mo oxo-thio complexes in 0.5 M H$_2$SO$_4$ under intermittent (1 sun) illumination. Applied potential = -0.1 V vs NHE.

Because low photoluminescence intensity in these materials is usually attributed to defect-mediated non-radiative recombination [33], the observed decreased recombination probability of photo-induced electrons and holes results from surface defects passivation by the thio-, oxo-thio-Mo-complexes.

To demonstrate that the recorded photocurrent could be ascribed to the hydrogen formation, we have measured the real-time hydrogen evolution by gas chromatography using a closed photo-electrochemical cell (See Supp. info). Under constant illumination, using a Pt calibration cell (Pt foils both as working and counter electrodes) and assuming a Faradic efficiency of 100 % for the Pt calibration device, we determined at j = 0.25 mA cm$^{-2}$ (E bias= - 0.1 V vs NHE) a Faradic yield around 97 %.

*3-4-Co-Catalyst Film Structure, Composition and Properties.*

To get a better insight into the co-catalyst - WSe$_2$ photoelectrode morphology, HRTEM was performed on the coated photo-electrodes samples. Characterization of samples before and after HER did not reveal significant changes in the film microstructure. Typical HRTEM images (Figure 3b) reveal an ultrathin amorphous film displaying irregular thickness ranging from 2 to 10 nm coating the well-crystallized 2D nanoflakes. The well-crystallized nanoflakes were unambiguously identified to WSe$_2$ from the lamellar structure displayed by the nanoflakes, in addition to the Se/W ratio value determined by EDS, (Se/W)$_{atom}$ = 2. Although an accurate determination of the S/Mo ratio is relatively tricky because of the overlap of the L$\alpha$ Mo (2.298 keV) and K$\alpha$ S (2.307 eV) energy peaks, S and Mo were nevertheless detected as major elements into the amorphous ultrathin film thus allowing full identification of the amorphous coating to the co-catalyst film. The low amount of the Mo thio oxo-thio catalyst deposited on the WSe$_2$ photo electrode revealed by HRTEM investigation was quantitatively assessed by Inductive Coupled Plasma measurements. After sonication and immersion of eight high performance electrodes (S/Mo= 5, pH 8.75, dip coated 2 x 16 h) in 10 ml NaOH 0.1 M solution, the loaded amount was measured to be 100 -150 Mo nanomoles cm$^{-2}$. With the assumption of a full coverage of the surface area of the 2D material including the basal plane and a head surface of 0.5 nm$^2$ for the Mo complexes, the average film thickness is thus determined in the order of some few layers (four to ten monolayers). These results suggest that when discussing on the catalytic performances of the film, we have to take into account properties of free-standing Mo complexes, while paying attention to the thio- and oxo- thio-Mo complexes anchored on the WSe$_2$ substrate.

Regarding the co-catalyst film composition, Fig. S7 shows the Raman spectrum recorded on high performance films prepared from solutions fabricated at S/Mo= 5 and in the pH range 8.75 < pH < 9.25. Although these films possess a S/Mo ratio = 5.0, Raman signatures including those previously recorded on isolated (Mo$_2$S$_{12}$)$^{2-}$ [14] and (Mo$_3$S$_{13}$)$^{2-}$ [13] clusters are detected in our films (Fig. S7) suggesting presence of these molecular structures into our amorphous films. Others molecular structures such as Mo monomers, and (Mo$_2$S$_6$O$_2$)$^{2-}$, (Mo$_3$S$_8$O)$^{2-}$ oxo-thio-Mo complexes present in the dip coating solution in peak concentrations cannot also be excluded as revealed from XPS spectra (Fig. 3a). To demonstrate the presence of these latter oxo thio Mo complexes in the dip coated solid film, fine inspection of the XPS data clearly reveals an average $r_{MoOS}$ ratio close to 0.23. Although this ratio was observed to be lower after complexes selection on dip coated $WSe_2$ electrodes compared to values determined on drop casted FTO ($r_{MoO}$ = 0.24-0.34) or on drop casted $WSe_2$ ($r_{MoO}$ = 0.36) electrodes, this reasonable high $r_{MoOS}$ ratio reveals that the complexes selection must include Mo oxo-thio-complexes such as $(MoOS_3)^{2-}$, $(Mo_2S_6O_2)^{2-}$ and $(Mo_3S_8O)^{2-}$ in addition to the Mo thio complexes. Another important point is the high value of $r_{S-ter}$, $r_{S-ter}$ > 0.35, detected by XPS on photo-catalytically efficient dip-coated $WSe_2$ electrodes (Fig. 3a). Although some terminal sulfide- disulfide groups could be partially transformed as a consequence of their high interaction with the $WSe_2$ photoelectrode substrate after anchoring on the substrate, the significant $r_{S-ter}$ value determined on our films indicate the presence of a high proportion of non-linked, sulfide-disulfide groups arising from high S/Mo ratio probably $(MoS_4)^{2-}$, $(Mo_2S_{12})^{2-}$ and $(Mo_2S_6O_2)^{2-}$ complexes, these latter complexes being either partially reticulated or included as isolated oligomers.

All these data indicate that our most efficient photocatalytic films are mainly composed of Mo monomers, $(Mo_2S_{12})^{2-}$, $(Mo_3S_{13})^{2-}$, $(Mo_2S_6O_2)^{2-}$ and $(Mo_3S_8O)^{2-}$ molecular structures. These building blocks could be present (i) in high interaction with the $WSe_2$ photo-electrode surface (ii) arranged into a partially reticulated network to form a continuous, mechanically stable film but with an important ratio of free terminal disulfide groups (iii) or as isolated oligomers as suggested by the high $r_{S-ter}$ value determined from XPS data.

Concerning the film formation, we propose that the co-catalyst film results from the conversion of the thio-, oxo-thio-Mo oligomers into an amorphous polymeric phase. Several mechanisms could be proposed: i) Formation of an oligomer-based coordination polymer involving interactions between the sulfide –disulfide groups and the Mo centres [30]. Indeed, the coordination-reticulation of the polymer is promoted by the various oxidation states displayed by the Mo centres as shown by XPS ii) oligomer-induced hetero-nucleation and growth as previously observed in the preparation of metal organic framework [46]. To account for the change of S/Mo ratio experimentally observed in our co catalyst films with dip coating time duration (S/Mo $_{atom}$= 4.2 -> 4.6), we propose that the first layers of thio-, oxo-thio-Mo oligomers anchored on the $WSe_2$ substrate serve to direct the formation of the subsequent layers by hetero-nucleation (or heterogeneous growth). The observed increase of S/Mo ratio thus arises from addition of selected oligomers displaying higher sulfide or disulfide groups density or higher S/Mo ratio such as $(Mo_2S_{12})^{2-}$. Lastly, these various interactions and reticulation are probably enhanced during the final drying stage improving the film consolidation.

Because our film forming process involves water-soluble Mo-complexes, a crucial advantage arising from the conversion of the Mo oxo-thio-oligomers into an amorphous solid film is its durability. This is illustrated on Fig. 4d showing a chrono-amperometric curve recorded under chopped light on a dip coated WSe$_2$ electrode. Moreover, XPS spectra recorded on films before and after the HER do not reveal significant change in S/Mo or $r_{S\text{-ter}}$ ratios. In accordance with previous works [13,14], our results demonstrate stability of the thio-Mo-complexes during the HER in the acidic electrolyte. Note that these first results are achieved on Mo complexes deposited on the 2D materials and we anticipate that this long term stability could be greatly improved by subsequent encapsulation in Nafion [14].

To provide insights into the performance enhancement resulting from the thio-, oxo-thio-Mo complexes film deposition, we further assess the PEC performances of the catalyst film and examine its electronic behaviour. Electrochemical Impedance Spectroscopy (EIS) measurements were performed on coated and non-coated WSe$_2$ photo-electrodes using a MoxSy catalyst film to provide an experimental evidence of the decrease of the charge transfer resistance resulting from the co-catalyst film deposition. Typical Nyquist plots are presented in Fig. S8a and S8b, respectively for bare and for coated WSe$_2$ electrodes under illumination at different potentials. One can notice that no important variation is observed for the bare electrode oppositely to the coated WSe$_2$, for which a decrease of the low frequency loop occurs with electrode polarisation. As a result, it is highly likely that the low frequency electrode impedance is driven by the charge transfer resistance [47] for the coated sample while charge recombination is the limiting step for the bare electrode. Because the different time constants were better highlighted in admittance plots compared with Nyquist representation, Fig. S8 c and d report the admittance plots recorded on bare and coated WSe$_2$ electrodes under illumination at different potentials. The admittance plots (Fig. S8c, S8d) of the bare and coated WSe$_2$ electrodes can be break down in three semi-circles. The one at high frequency (right side of the plot) exhibits constant diameter over the polarisation, the two others ones (from right to left) show variable diameters for the coated samples but not for the bare one. Indeed, two loops are clearly visible particularly at low polarisation, –beyond + 0 mV vs NHE- until only one prevails at higher polarisation.

To get a better insight regarding the charge process within the different electrodes, an equivalent circuit has thus been proposed taking into account the different time constants. We propose to use a string of three $Z_{arc}$ (see details in Fig. S8e,f, Supp. Info) in series with series resistance ($R_S$), the latter standing for the electrolyte resistance. Each $Z_{arc}$ has been ascribed to different processes $Z_{arc}(1)$ for the so called bulk impedance, -the electric impedance of the electrode- $Z_{arc}(2)$ for the impedance relative to the surface states [47] or recombination process and $Z_{arc}(3)$ for the interface impedance (charge transfer resistance together with the interface capacitance). In Table S1 and S2 (see Supp. inf.) are reported the different values obtained. One can notice for coated sample that beyond -0 mV NHE- only one time constant has been needed to fit the curves confirming what was observed in the respective admittance plots. The interface capacitance values extracted for the bare and coated WSe$_2$ electrodes are reported Fig. 5a showing no significant changes. Because these capacitance values give a good estimation of the electrochemical surface area, the nearly constant values of the interface capacitance determined for both electrodes indicate that the impedance of the interface is mainly driven by charge transfer and recombination resistances. From these data it has been extracted two kinds of resistance, one is ascribed to recombination processes, the other to charge transfer (Fig. 5b). As described in Supp. Info., a series equivalent circuit have been chosen meaning that the limiting step is dictated by the higher resistance. In this model, the more there is charge carriers recombination, the higher will be the resistance, denoted $R_{rec}$. As it can be noticed, the coated $WSe_2$ electrodes display overall resistance values – the sum of the two resistances – always lower than those determined for the bare $WSe_2$ electrodes. Having a closer look to the respective resistances, one can notice the $R_{rec}$ of the coated $WSe_2$ is always one order of magnitude below the charge transfer resistance one which makes the latter drives the electrode kinetic. For the bare $WSe_2$ electrode, the two resistance values are closer and could explain why the impedance of this electrode is less sensitive to the electrode polarisation since the recombination is a non-negligible part of the whole electrode process. As a result, all our observations suggest that the limiting step determined for the bare electrode would be charge recombination while predominance of charge transfer for the coated electrode strongly indicates surface defect passivation by the co-catalyst film.

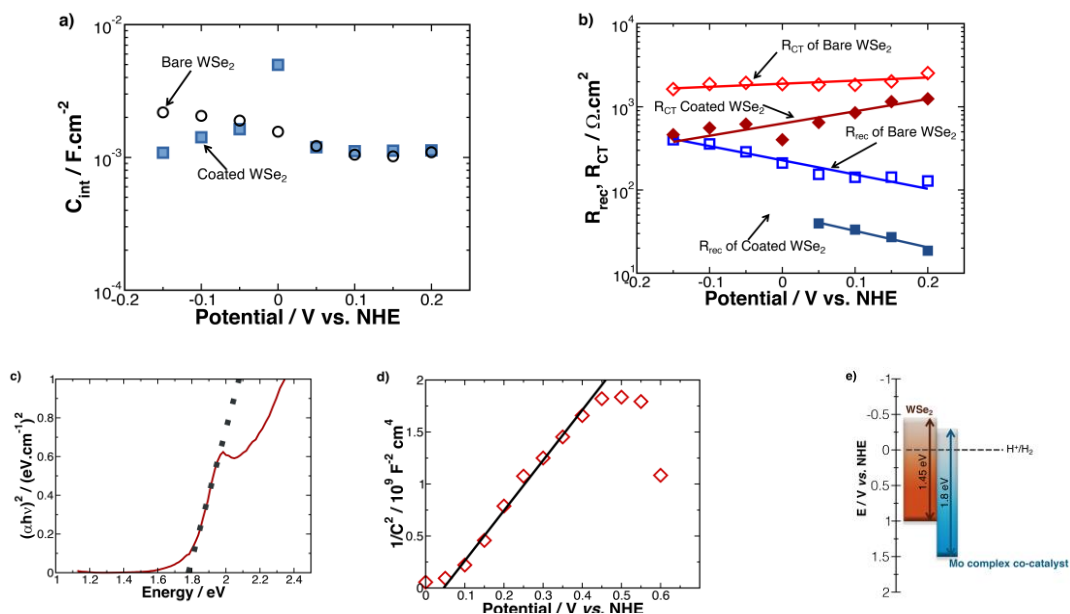

**Fig. 5.** Optical and photoelectrochemical properties of co catalyst -WSe$_2$ photoelectrodes. a) Interfacial capacitances and b) Charge transfer and recombination resistances extracted from impedance curves recorded at various applied potentials and under illumination (λ= 450 nm) for bare WSe$_2$ electrode and co-catalyst coated WSe$_2$ electrode. Co-catalyst film = Thio, oxo-thio-Mo complexes pH 9, S/Mo= 5. Electrolyte 0.5 M H$_2$SO$_4$, pH 1. c) UV-Vis absorption curve recorded on Mo complexe co-catalyst film drop casted on FTO substrate. d) Mott-Schottky plots of Mo complexe co-catalyst –FTO showing a positive slope and a flat band potential $E_{Fb}$= +0.05 V vs NHE at pH 7. The co-catalyst film was prepared by drop casting, air evaporation and heat treated in air atmosphere at 110 °C, 10 min. Electrolyte 0.1M Na$_2$SO$_4$, pH 7. Capacitances values were extracted from the EIS measurements at 100 Hz. e) Schematic for the energy band structure of Mo complexe co-catalyst-WSe$_2$ heterojunction. The band gap value of WSe$_2$ was determined from transmittance-reflectance curves recorded on WSe$_2$ nanoflakes film drop casted on a glass substrate.

To better ascertain the functionalities displayed by the various layers building up the photocathode, the electronic properties of the co-catalyst film were next assessed. The semiconducting property of the catalyst film was revealed from UV-Vis absorption spectra (Fig. S9) recorded on co-catalyst film deposited on glass substrates and annealed at 110 °C. The direct optical transition of the co-catalyst film was demonstrated from the straight line observed by plotting *($\alpha h\nu)^2$* against photon energy (*$h\nu$*) (Fig. 5b) from which a band gap of 1.80 eV can be extrapolated. Moreover, properties of the semiconducting catalyst film-electrolyte interface were investigated in 0.1 M $Na_2SO_4$, pH 7, using a co-catalyst film deposited on a FTO-Glass substrate. The Mott-Schottky plots (Fig. 5c) determined from the capacitance measurements revealed a n-type semi-conduction, a flat band potential of -0.20 V vs NHE (pH 1) and a charge carrier density of $10^{-21}$ $cm^{-3}$. From this donor density along with the flat band potential value, position of the conduction band was determined at -0.3 V vs NHE (For details, See section 3, Supp. Info.).

To construct the band energy diagram of thio-, oxo-thio Mo complexes co catalyst film and $WSe_2$, we have determined the band gap of p-$WSe_2$ films formed from our exfoliated nanoflakes (Eg = 1.45 eV). Using the previous optoelectronic properties determined for the co-catalyst film, and a conduction band energy level of -0.45 V vs NHE [48] for the $WSe_2$ nanoflakes film, the band energy diagram (Fig. 5d) clearly shows that the n-$Mo_xS_y$ co-catalyst film is suitable for ensuring efficient charge separation and migration of the photo-induced electrons from the p-$WSe_2$ photo-electrode to produce hydrogen.

In addition to develop a high catalytic activity, the multicomponent co-catalyst film was shown to exhibit multiple advantages. The Mo complexes co-catalyst film possesses a n-type semi-conduction yielding a p/n junction resulting in increased band bending [49]. More interestingly, the co-catalyst film is able to heal photo-electrodes displaying high surface defects density such as 2D WSe$_2$ nanoflakes. Although the multicomponent character make complex the film composition and structure, respective roles of the various complexes as healing additives and catalyst were explored on model Mo thio- complexes and supported by DFT calculations.

*3-5-Insights into the Respective Roles of Mo Monomers and Mo Thio-, Oxo-Thio Complexes as Healing Additives and Catalysts.*

Fig. S10 reports typical J-V curves recorded on films prepared by selective dip coating in Mo complexes solutions fabricated at S/Mo= 5 but prepared at various pH or impregnation time durations. Depending on their forming conditions, films displaying similar dark current (or electro-catalytic activity) can exhibit different photocurrent densities ranging from 0.25 to 2.0 mA cm$^{-2}$. On the other hand, films exhibiting different electro-catalytic activities can display a similar photocurrent up to ~ 2 mA cm$^{-2}$. These non-correlated evolutions of catalytic activity and photocurrent density thus demonstrate that our films display a multicomponent character and are composed of various Mo complexes with specific properties. These results suggest that our films do not act as simple protective layers [50,51] such as previously described ultrathin Al$_2$O$_3$ films [50]. More particularly, our films are formed from a mixture of Mo complexes displaying high catalytic activity with complexes capable to heal surface defects.

Because our Mo complexes are prepared in alkaline conditions where monomers and oxo-thio-complexes are in significant proportions, it is thus crucial when designing all-aqueous Mo complexes to better evaluate the respective contributions of the thio Mo complexes ((MoS$_4$)$^{2-}$, (Mo$_2$S$_{12}$)$^{2-}$) , also, of the oxo-thio Mo complexes on the catalytic activity and the healing effect. Properties of WSe$_2$ electrodes coated exclusively with pure (Mo$_2$S$_{12}$)$^{2-}$ prepared in DMF as previously described in the literature [14] were thus investigated in a large range of catalyst loadings (30 < n nanomoles < 500). Large current densities up to 10 mA cm$^{-2}$ at -0.4 V vs NHE in the range of the best current densities recorded on films prepared from our water-soluble thio Mo complexes were achieved with pure (Mo$_2$S$_{12}$)$^{2-}$ highlighting the high electro-catalytic activity displayed by this complex. Concerning its photo-catalytic properties, the photocurrents recorded on pure (Mo$_2$S$_{12}$)$^{2-}$ films remain significantly much lower than those observed on films fabricated by dip coating from our oxo-thio Mo complexes solutions (Fig. S10b). Therefore, the higher photocurrent densities up to 2 mA cm$^{-2}$ at -0.2 V vs NHE recorded on our films highlight the beneficial effect of the Mo thio monomers or oxo-thio complexes compared with pure (Mo$_2$S$_{12}$)$^{2-}$ to heal the 2D WSe$_2$ surface defects.

To support these findings, DFT calculations of adsorption energies and catalytic activity of the various Mo complexes were performed. As previously discussed, the various molecular structures can occur either as a reticulated polymer or as isolated building blocks. Moreover, no significant evolution was observed on the Raman spectra recorded before and after acid soaking or HER cycling (Fig. S7) revealing no changes of structure or composition of the thio-, oxo-thio-Mo complexes during the HER in acidic solutions consistent with previously results reported in the literature [13,14]. From all these results concerning the film composition, as a primary screen, we have investigated the roles of free-standing as well as anchored thio-, oxo-thio- Mo complexes respectively as healing additives and catalysts. More particularly, our DFT calculations were performed on thio, $(MoS_4)^{2-}$, oxo-thio, $(MoS_3O)^{2-}$ monomers and $(Mo_2S_{12})^{2-}$, $(Mo_2S_6O_2)^{2-}$ dimers.

Catalytic Activities of free-standing $(MoS_3O)^{2-}$, $(Mo_2S_{12})^{2-}$ and $(Mo_2S_6O_2)^{2-}$ Complexes.
More insights into the intrinsic catalytic activities of the $(MoS_4)^{2-}$, $(MoS_3O)^{2-}$ monomers and $(Mo_2S_{12})^{2-}$, $(Mo_2S_6O_2)^{2-}$ dimers were gained in a first stage by gas-phase DFT calculations. This was carried out using a molecular approach (For details, see Section 2, Supp. Info.) on H absorption energies on free-standing $(MoS_4)^{2-}$, $(MoS_3O)^{2-}$ monomers and free-standing $(Mo_2S_{12})^{2-}$, $(Mo_2S_6O_2)^{2-}$ dimers in their radical forms. We recall that the free Gibbs energy of H-adsorption is a good descriptor of the ability of a given compound to be active for HER, when $\Delta G_H$ remains close to ± 0.1 eV [34]. Unlike $(MoS_4)^{2-}$ and $(MoS_3O)^{2-}$ monomers, low hydrogen adsorption energies were determined on free-standing $(Mo_2S_{12})^{2-}$ and $(Mo_2S_6O_2)^{2-}$ complexes (Fig. 6). While confirming the low hydrogen adsorption energy of the $(Mo_2S_{12})^{2-}$ thio complexes reported in literature [14], these calculations highlight the ability of oxo-thio Mo complexes such as $(Mo_2S_6O_2)^{2-}$ to catalyse the HER. In addition, it is worthy to note that from our DFT calculations the M=O units were identified as the more active site for the HER on $(Mo_2S_6O_2)^{2-}$ complexes.

Adsorption energies of thio-, oxo-thio-Mo complexes on point defects of 2D $WSe_2$ materials.
Defects in 2D TMDC nanosheets occur on the edges or are located in the basal plane due to non-stoichiometry. From TEM observations, our $WSe_2$ nanoflakes possess average lateral dimensions of 1 µm x 1 µm and a thickness of about 5 monolayers. When assuming a defect concentration of 0.5 % for the peripheral in-plane sites, a simple calculation reveals an equivalent number of peripheral edge or in-plane defect sites. In addition, and as expected, preliminary DFT calculations have shown a stronger adsorption of the thio complexes on the edge sites. From these results, we have paid particular attention in our DFT calculations to point defects. Regarding point defects, best performance for the HER was achieved on films prepared from solutions containing $MoS_4^{2-}$, $MoS_3O^{2-}$ monomers, $(Mo_2S_{12})^{2-}$ and $(Mo_2S_6O_2)^{2-}$ dimers in high proportions. During the catalyst film formation, these negatively charged thio-complexes preferably interact electrostatically in basic aqueous solution with locally charged defects of the 2D substrate. Reported defects on the 2D substrate include edges [6], large scale defects such as tears [32,52] or point defects such as Se vacancies, W vacancies [53]. Typical point defects of the WSe$_2$ substrate, namely Se-vacancy and W-vacancy are known to be electron acceptors [53,54] with formation energy we have determined respectively of 2.6 and 5.3 eV. Note that these vacancies always yield very small atomic reorganization with stretching of the W-Se bonds less than 0.1 Å (Fig. S11). In addition to have a considerable effect on carrier recombination, scaling of these point defects with the molecular size of the Mo complexes makes a meaningful DFT investigation focussed on point defects rather than large scale defects.

| Adsorption energies onto WSe$_2$ (in eV) | Free –defect Pristine substrate | Se-Vacancy containing WSe$_2$ substrate | W-vacancy containing WSe$_2$ substrate |
|---|---|---|---|
| (MoS$_4$)$^{2-}$ | +0.09 | -0.15 | -1.75 |
| (MoS$_3$O)$^{2-}$ | -0.59 | -0.97 | -2.64 |
| (Mo$_2$S$_{12}$)$^{2-}$ | -0.09 | -0.11 | -0.29 |
| (Mo$_2$S$_6$O$_2$)$^{2-}$ | -0.13 | -0.19 | -0.99 |

**Table 1**. Adsorption energies (in eV) of various Mo-complexes in their radical forms, for three distinct substrate configurations of the WSe$_2$ monolayer.

The Mo complexes, (MoS$_4$)$^{2-}$, (MoOS$_3$)$^{2-}$ monomers, and (Mo$_2$S$_{12}$)$^{2-}$, (Mo$_2$S$_6$O$_2$)$^{2-}$ dimers chosen as model Mo species were put in interaction in their radical form with (i) a pristine, (ii) a single Se-vacancy and (iii) a W-vacancy WSe$_2$ monolayer. Table 1 summarizes the adsorption energy of Mo-complexes, defined as E$_{ads}$= E$_{Mo-comp@subst}$ - E$_{subst}$ -E$_{Mo-comp}$ on a model substrate presenting (or not) Se or W vacancy, where E$_{Mo-comp@subst}$, E$_{subst}$, and E$_{Mo-comp}$ stand for the total energy of the adsorbed species on the model WSe$_2$ substrate, for the total energy of the substrate alone and of the Mo-complex alone respectively. From the values, it appears that the Mo-complexes weakly bind with defect-free pristine WSe$_2$ substrate and always prefer to stack on W-vacancy (see Fig. 6 for the adsorption modes on the W-vacancy site and Fig. S12a,b for other adsorption sites). Interestingly, (MoS$_4$)$^{2-}$, (MoS$_3$O)$^{2-}$ monomers or polycondensed species displaying a small size such as (Mo$_2$S$_6$O$_2$)$^{2-}$, were shown to bind stronger than the (Mo$_2$S$_{12}$)$^{2-}$ complex to point defects of the WSe$_2$ substrate. Note that DFT calculations show that oxo-thio complexes are better adsorbed compared with their thio analogues on the WSe$_2$ substrate containing Se or W vacancies. In contrast to the disruption of S$_2$$^{2-}$ bonds into active S$^{2-}$ anions previously reported in the presence of nucleophiles favouring subsequent reactions with metal ions producing metal sulphides [56,57], the interaction between a W-vacancy and the various complexes investigated (($MoOS_3$)$^{2-}$, ($Mo_2S_6O_2$)$^{2-}$, ($Mo_2S_{12}$)$^{2-}$) results in the formation of new Se-S bonds (d $_{Se-S}$ = 2.3 Å) as illustrated on Fig. 6a. Therefore, we propose that the passivation of the surface defects of the $WSe_2$ nanosheets arises from the strong adsorption of these Mo complexes onto the $WSe_2$ nanosheets while interacting with the Se and W vacancies. Note that the stronger adsorption displayed by Mo monomers or small size dimers such as ($Mo_2S_6O_2$)$^{2-}$ toward point defects shown from our calculations is consistent with our experimental results. This highlights the crucial role of these latter Mo complexes compared with pure $(Mo_2S_{12})^{2-}$ to passivate the surface defects on 2D $WSe_2$.

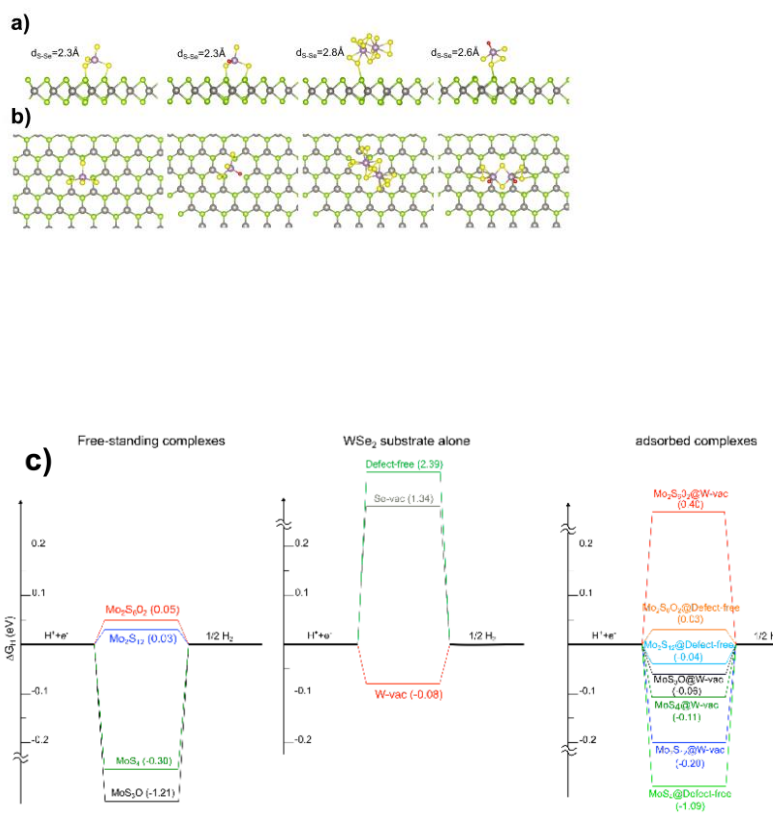

**Fig. 6.** Side views (a) and top views (b) of the atomic structures for the Mo oxo-thio complexes on $WSe_2$ monolayer possessing a W-vacancy, showing S…Se interactions and accessible disulfide groups for the dimer Mo complexes. From left to right: adsorption of $MoS_4$, $MoS_3O$, $Mo_2S_{12}$ and $Mo_2S_6O_2$ radicals. Yellow: Sulfur, Red: Oxygen, Violet: Mo, Green: Se and Grey: W. (c) Free energy diagram for Hydrogen evolution at equilibrium. ($c_1$) $\Delta G_{ads}H$ showing values close to the optimum of the volcano curve for free standing ($Mo_2S_{12}$)$^{2-}$ and ($Mo_2S_6O_2$)$^{2-}$ complexes ($c_2$) $\Delta G_{ads}H$ values for $WSe_2$ monolayers alone, namely point defect –free or W or Se vacancies-containing $WSe_2$ monolayers ($c_3$) $\Delta G_{ads}H$ values for various adsorbed Mo complexes on point defect –free or W or Se vacancies-containing $WSe_2$ monolayers.

Catalytic Activities of Thio-, Oxo-Thio-Mo Complexes in Anchored Configuration.

Although in our DFT calculations we exclusively consider point defects and deal with adsorption of radical vs charged Mo complexes, our calculations highlight that good catalytic performance will mainly arises from dimers such as $((Mo_2S_6O_2)^{2-}$ or $(Mo_2S_{12})^{2-}$ preferably in a free standing regime or anchored in a low adsorption regime, i.e. on point defect-free $WSe_2$ substrates (Fig. 6c). In contrast, $(MoS_4)^{2-}$, $(Mo_2S_6O_2)^{2-}$, $(Mo_2S_{12})^{2-}$ complexes strongly adsorbed on W-vacancy containing $WSe_2$ monolayers show $\Delta G_{ads}H$ values far from the optimum of the Volcano curve [34] indicating a low activity for the HER. Thus, DFT calculations showed that our water-soluble Mo-oxo-thio complexes distributions are composed of three various classes of molecular species: i) Mo monomers possessing a low catalytic activity in its free-standing configuration but displaying high adsorption energy onto 2D $WSe_2$. ii) $(Mo_2S_{12})^{2-}$ dimer displaying high catalytic activity for the HER but with relatively low affinity for the 2D $WSe_2$ substrate. iii) $(MoS_3O)^{2-}$ oxo thio monomer displaying strong adsorption towards the $WSe_2$ substrates and reasonable catalytic activity in its anchored configuration. These Mo oxo-thio complexes distributions are capable both of passivating surface defects and displaying catalytic activity can therefore be described as healing catalysts.

Consistent with our experimental results, we propose that the observed high HER performance arises from oxo-thio Mo complexes films probably composed of Mo monomers, dimers and trimers which are present in addition to a reticulated polymer both in free-standing and anchored configurations. Low adsorption and high adsorption modes of the Mo complexes should coexist for the anchored configurations depending of the point defects local concentration. We have demonstrated that the optimum structure of the catalyst film including free-standing and strongly adsorbed complexes providing respectively catalytic and healing properties can be successfully achieved from well-defined aquo-oxo-thio Mo complexes distributions through selective dip coating. Nevertheless, deep insights into the film structure describing respective proportions of healing and catalytic additives as well as the detailed conformation of the polymer and the complexes inside the film remain to be elucidated.

*3-6-Implications for the Hydrogen Evolution Reaction Mechanism.*
Different scenarios are proposed to date for the Hydrogen Evolution Reaction catalytic cycles on Mo thio complexes especially in the presence of M=O moiety [26]. More particularly, the HER catalytic cycle reported in the case of high performance amorphous MoSx films prepared by electrodeposition involves unsaturated Mo(IV) and Mo(V) sites with transient formation of the M=O moiety during the catalytic cycle. In another study concerning the same material, the catalytic cycle is shown to involve the $S_2^{2-}$ -> $S^{2-}$ redox reaction [26]. In recent works [25,26] dealing on MoO(S$_2$)$_2$(2,2'-bipyridine) as oxo thio Mo model molecule, a H$_2$ evolution catalytic cycle involving a hardly attached M=O moiety which functions as a proton relay is proposed for the HER reaction. In this catalytic mechanism, the $S_2^{2-}$ anion represents the redox centre associated with a Mo(V) site. Note that these various catalytic cycles were proposed for the hydrogen electrocatalytic evolution reaction. A crucial difference between the photocatalytic vs electrocatalytic electrochemical water decomposition, is the shift of potential arising from the solar energy harvesting. This shift towards lesser cathodic potentials does not favour the Mo(V)-> Mo(IV) reduction which is reported to occur at -0.3 V /NHE in slightly acid medium [30]. Because hydrogen evolution in our photocatalytic experiments does not require any cathodic pre-activation nor the high cathodic potentials for the Mo(V)-> Mo(IV) reduction, we believe that our photocatalytic cycle for the HER implies terminal $S_2^{2-}$.

This work should pave the way to further design a larger range of environmentally-friendly, aqueous-stable, multicomponent [58], catalytically active passivation layers made of thio-, oxo-thio- Mo complexes. Besides developing catalytic activity and healing property, these multicomponent films with a n-type semi-conduction along with its suitable optoelectronic properties were shown to improve charge separation and migration for the HER thus providing an example of engineering of multicomponent, passivation layer displaying multiple properties.

While this work was focused on 2D WSe$_2$ materials which possess a considerable surface defect density, use of these highly interactive, oxo-thio, molecular complexes could be extended to passivate surface defects of a larger range of high surface area photo-electrodes like p-WS$_2$ or p-MoS$_2$ for the HER. Further developments could also be anticipated with the design of M doped- Mo-oxo-thio-complexes [59], where M= metallic cation [24], rare earth elements [60], especially to achieved lower cathodic HER onset potentials or higher catalytic activity in neutral aqueous electrolytes. Indeed, because of their ability to absorb a significant fraction of the solar spectrum using ultrathin films, 2D TMDC materials (p-MoS$_2$, p-WS$_2$, p-WSe$_2$) displaying various band gaps and onset potentials for HER [48] represent ideal photocathode candidates for the engineering of photo-electro-chemical tandem cells [61].

**4-Conclusions**

Co-catalyst films composed of Mo thio- and oxo-thio- complexes which spontaneously formed in H$_2$O in well-defined S/Mo ratio and pH conditions, are proposed as HER catalysts.

Photocatalytic decomposition of water was successfully achieved on photocathodes prepared from exfoliated 2D-$WSe_2$ and activated by highly interacting complexes selected from Mo oxo-thio complexes distributions via a selective dip coating film forming process. Best photocatalytic results were observed on 2D p-$WSe_2$ photo-electrodes after deposition of Mo monomers, $(Mo_2S_{12})^{2-}$, $(Mo_2S_6O_2)^{2-}$ and $(Mo_3S_8O_2)^{2-}$ with photocurrents up to 2.0 mA cm$^{-2}$ at - 0.2 V / NHE. Compared with Pt-Cu catalysts, the higher photocurrents observed after deposition of the thio- and oxo-thio- Mo complexes, without the use of any passivating additive on the 2D photo-electrodes, reveal a healing effect arising from the oxo-thio Mo complexes films. The healing property of the co-catalyst film was experimentally demonstrated by photoluminescence and electrochemical impedance spectroscopy. Insights into adsorption energies and final conformations of the thio and oxo-thio complexes on the 2D photo-electrodes are given from DFT calculations. DFT results highlight the crucial healing role of oxo-thio Mo monomers and to a lesser extent, of smaller oxo- thio- Mo dimers. These films also provide an example of the engineering of a multicomponent passivation layer displaying multiple properties including healing, better charge separation and migration, catalytic activity. The strategy illustrated here on 2D materials and relying on strongly adsorbed molecular species could be extended to other large surface area photo-electrode materials displaying high concentrations of surface defects.

**Methods.**

The experimental details can be seen in the Supporting Information.

**Acknowledgements.**

J. B. B. and L. G. F. V. thank Conselho Nacional de Desenvolvimente Cientifico e Technologico (CNPQ), Brazil, for the financial support through CNPq grants (CNPq 201490/2015-3 and INAMI/CNPq/MCT). I. T. M. is grateful for the computational resources provided by the high-performance computational facility of the Babes-Bolyai University (MADECIP, POSCCE,COD SMIS 48801/1862) co-financed by the European Regional Development Fund of the European Union. I. C. G. and R. P. thank the CALMIP for their generous allocation of computational resources (N° p0812). This work was also granted access to the HPC resources of CINES and IDRIS under the allocation 2018-A0040906649 made by GENCI.

**Conflict of Interest**

The authors declare no conflict of interest.

**Supporting information.**

Experimental section. Peaks identification from ESI-MS data. Additional XPS, Raman and photo electrochemical data. Impedance plots. Additional DFT results.